\documentclass[smallextended]{svjour3}     
\usepackage[utf8]{inputenc}
\usepackage{braket}
\usepackage{float}
\usepackage{amsmath}
\usepackage{tikz}
\usepackage{graphicx}
\usepackage{caption}
\usepackage{subcaption}
\usepackage[english]{babel}
\usepackage{dirtytalk}
\usepackage{amsmath} 
\usepackage[utf8]{inputenc}
\usepackage[none]{hyphenat}
\usepackage[colorlinks=true, citecolor=blue]{hyperref}
\usepackage{mathtools}
\usepackage{bm}
\usepackage{amssymb}
\usepackage[a4paper,left=17mm,top=22mm,right=17mm,bottom=15mm]{geometry}
\smartqed

\begin{document}

\title{Generalization and Demonstration of an Entanglement Based  Deutsch-Jozsa Like Algorithm Using a 5-Qubit Quantum Computer}

\titlerunning{Generalization ... of an Entanglement Based  Deutsch-Jozsa Like Algorithm Using a 5-Qubit Quantum Computer}        

\author{Sayan Gangopadhyay$^\P$         \and
        Manabputra$^\P$                  \and
        Bikash K. Behera                 \and
        Prasanta K. Panigrahi
}

\authorrunning{Sayan Gangopadhyay \& Manabputra et al.}

\institute{$^\P$These authors have contributed equally to this paper\\
\newline
Sayan Gangopadhyay$^\P$ \at
              Undergraduate Programme, Indian Institute of Science, Bangalore 560012, India \\
              \email{sayangangopadhyay48@gmail.com}           
           \and
           Manabputra$^\P$ \at
             School of Physical Sciences, National Institute of Science Education and Research, HBNI, Jatni 752050, Odisha, India \\
              \email{manabputra@gmail.com}
              \and
              Bikash K. Behera \at
              Department of Physical Sciences, Indian Institute of Science Education and Research Kolkata, Mohanpur 741246, West Bengal, India \\
              \email{bkb13ms061@iiserkol.ac.in}
              \and
              Prasanta K. Panigrahi \at
              Department of Physical Sciences, Indian Institute of Science Education and Research Kolkata, Mohanpur 741246, West Bengal, India \\
              \email{pprasanta@iiserkol.ac.in}
              }

\date{Received: 11 January 2018 / Accepted: date}

\maketitle
\begin{abstract}
This paper demonstrates the use of entanglement resources in quantum speedup by presenting an algorithm which is the generalization of an algorithm proposed by Goswami and Panigrahi [arXiv:1706.09489 (2017)]. We generalize the algorithm and show that it provides  deterministic solutions having an advantage over classical algorithm. The algorithm answers the question of whether a given function is constant or balanced and whether two functions are equal or unequal. Finally, we experimentally verify the algorithm by using IBM's five-qubit quantum computer with a high fidelity.
\end{abstract}
\section{Introduction}
\label{qal_intro}
Deutsch-Jozsa algorithm \cite{qal_Jozsa} is one of the first examples of quantum algorithm which is exponentially faster than any possible deterministic classical algorithm that solves the same problem. A special case of Deutsch-Jozsa problem is Deutsch problem \cite{qal_Deutsch}; given that a function $f:\big\{0, 1\big\} \rightarrow \big\{0, 1\big\}$; is either constant or balanced, the task is to determine whether the function is constant or balanced. Goswami and Panigrahi recently provided a quantum algorithm \cite{qal_Ashutosh} which uses entanglement as a resource for quantum speedup. They have shown for two black boxes $f$ and $g$, with a promise that either both are constant or both are balanced, one can solve the following two problems in a single use of each function:  
    \begin{itemize}
        \item They are constant or balanced. 
        \item They are equal or unequal. 
     \end{itemize}
     
In this paper, we consider the generalization of this algorithm, where we take $n$ black boxes instead of two. We show explicitly how entanglement helps in reducing the number of queries required to solve the aforementioned problem. We show that our algorithm gives deterministic result and reduces the number of queries by one. \par

Recently, IBM has developed a 5-qubit quantum processor, ibmqx4, which is the world's first commercial quantum computing service provided by IBM via a free web based interface called \textit{IBM Quantum Experience} (IBM QE) \cite{qal_IBM}. Researchers have taken a proper advantage of it by demonstrating and running a variety of quantum computing experiments, e.g., \cite{qal_LeggetIBM,qal_MerminIBM,qal_EntropicIBM,qal_CloudIBM,qal_LinkeIBM,qal_WoottonIBM,qal_BKB1,qal_MS1,qal_MS2,qal_MS3,qal_AK1,qal_BKB2,qal_BKB3,qal_BKB5}. Hence, we have implemented different cases of the proposed algorithm using the IBM quantum computer. 

The rest of the paper is organized as follows. Section \ref{qal_II} describes some preliminary concepts about Deutsch-Jozsa Algorithm. Section \ref{qal_III} proposes our algorithm, following which Section \ref{qal_V} demonstrates the experimental verification of the algorithm through IBM quantum experience. Finally, Section \ref{qal_VI} concludes the paper by summarizing as well as providing future directions of our work.

\section{Deutsch-Jozsa Algorithm: Some Preliminaries \label{qal_II}}

We have a function $f:\big\{0, 1\big\}^n \rightarrow \big\{0, 1\big\},$ where $n\in\mathbb{N}$; such that $f$ satisfies one of the following two possibilities.
\begin{itemize}
    \item $f$ is constant i.e, $f(x)=0$ or $f(x)=1 $, $ \forall  x \in \big\{0, 1\big\}^n$
    \item $f$ is balanced i.e, $|\big\{x \in\big\{0, 1\big\}^n  : f(x)=0 \big\}|=|\big\{x \in\big\{0, 1\big\}^n  : f(x)=1 \big\}|=2^{n-1}$
\end{itemize}
Here, the task is to determine which of the two above properties, is satisfied by $f$. Best deterministic classical algorithm takes $2^{n-1}-1$ queries to solve this task since one may get $2^{n-1}$ $0's$ before getting a 1 \cite{qal_nil}, whereas Deutsch-Jozsa Algorithm solves this problem in only $1$ query. We assume that the function $f$ is calculated using unitary $U_f :\Ket{x}\Ket{y}\rightarrow\Ket{x}\Ket{y\oplus f(x)}$.
The Deutsch-Jozsa Algorithm has been explicated in the following circuit.

\begin{figure}[H]
\includegraphics[scale=1]{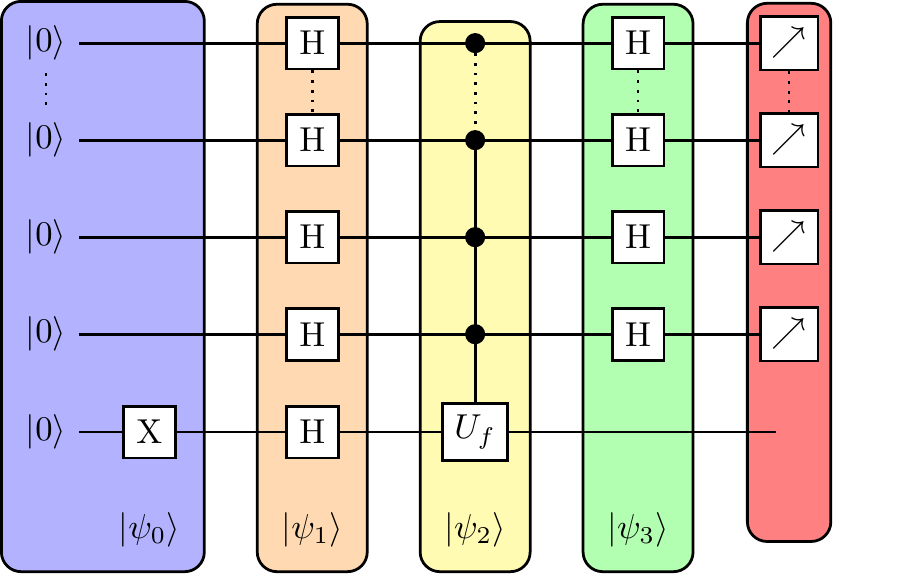}
\caption{Circuit implementing the general Deutsch–Jozsa Algorithm.}
\label{qal_fig1}
\end{figure}
 The input state $\Ket{\psi_0}=\Ket{0}^{\otimes n}\Ket{1}$, after application of Hadamard gates we get,
 \begin{align}
  \Ket{\psi_1}=\sum_{x \in\big\{0, 1\big\}^n} \frac{\Ket{x}}{\sqrt{2^n}}\Bigg[\frac{\Ket{0}-\Ket{1}}{\sqrt{2}}\Bigg]    
 \end{align}
In Figure \ref{qal_fig1}, the first n wires represent n-qubit query register and the last wire represents the answer register. Now the function is evaluated using unitary $U_f$ operations on query register as well as on answer register. The obtained state,
  \begin{align}
 \Ket{\psi_2}=\sum_{x \in\big\{0, 1\big\}^n} \frac{(-1)^{f(x)}\Ket{x}}{\sqrt{2^n}}\Bigg[\frac{\Ket{0}-\Ket{1}}{\sqrt{2}}\Bigg]
 \end{align}
 contains the result of the evaluation of a function in a superposed state. Eventually, application of Hadamard gates to query registers reveals whether $f$ is constant or balanced as described below.
 
 \begin{align}
 \Ket{\psi_2}\overset{H^{\otimes n} \otimes I}{\longrightarrow }\Ket{\psi_3}=\sum_{x \in\big\{0, 1\big\}^n}\sum_{z \in\big\{0, 1\big\}^n} \frac{(-1)^{x.z + f(x)}\Ket{x}}{2^n}\Bigg[\frac{\Ket{0}-\Ket{1}}{\sqrt{2}}\Bigg]
 \end{align}
 when $z=\ket{0}^{\otimes n}$ and $f$ is constant,
 \begin{align}
 \Ket{\psi_3}=\sum_{x \in\big\{0, 1\big\}^n} \frac{(-1)^{f(x)}}{2^n}\Ket{x}\Bigg[\frac{\Ket{0}-\Ket{1}}{\sqrt{2}}\Bigg]
 \end{align}
 Hence, the probability of obtaining $\ket{0}^{\otimes n}$ in query register is given as, 
 \begin{align}
  \Bigg|\sum_{x \in\big\{0, 1\big\}^n}\frac{(-1)^{f(x)}}{2^n}\Bigg|^2=
 \begin{cases}
     1, & \mbox{ if $f$ is constant}\\
     0, & \mbox{ if $f$ is balanced}\\
   \end{cases}
\end{align}
It is found that, if one gets $|0\rangle$ by measuring each of the qubits in the query register, then the function is constant otherwise balanced.\par

\section{The Proposed Algorithm \label{qal_III}}

\textbf{Problem:} Given that there are $n$ functions, $f_i:\{0,1\}\rightarrow \{0,1\}$; such that either $f_i$ is constant or balanced $\forall i \in [1,n]$. One needs to determine whether the functions are constant or balanced and whether they are equal or unequal (even if one function is different from the rest, the conclusion should be `unequal').

\begin{figure}[H]
\includegraphics[scale=0.9]{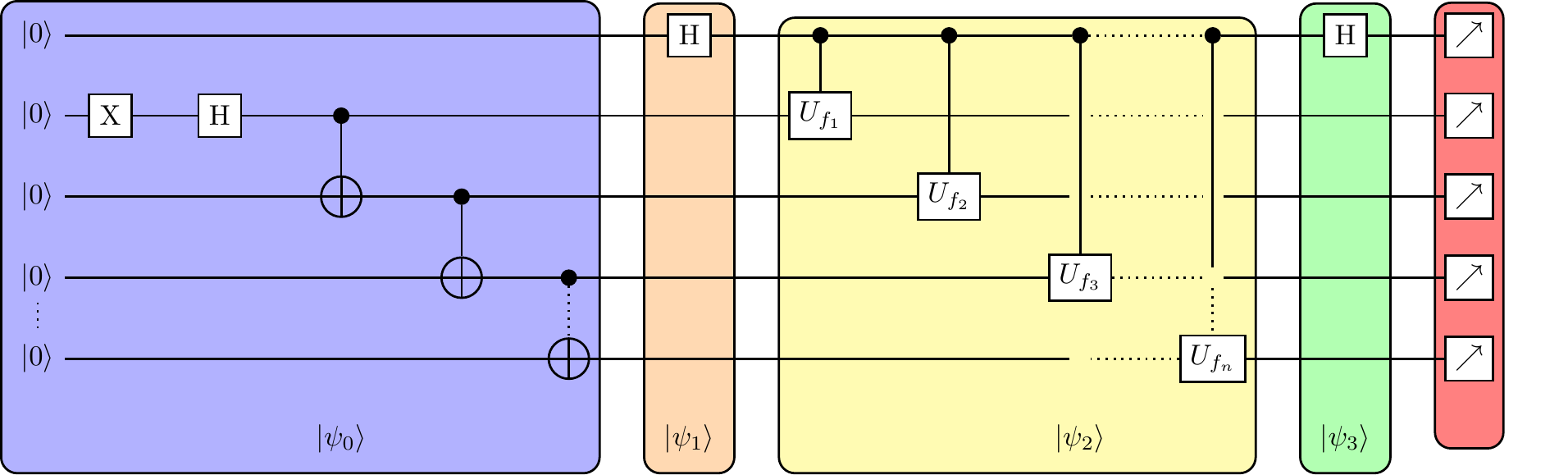}
\caption{Circuit depicting the proposed Algorithm. The first qubit signifies the register qubit and the rest of the qubits represent answer qubits.}
\end{figure}

\par Suppose, there are $n$ functions and $n$ entangled qubits of the form $\frac{\ket{0}^{\otimes n}-\Ket{1}^{\otimes n}}{\sqrt{2}}$, then the initial state, 
\begin{align}
\Ket{\psi_0}=\Ket{0}\Bigg\{\frac{\Ket{0}^{\otimes n}-\Ket{1}^{\otimes n}}{\sqrt{2}}\Bigg\}
\end{align}
\hspace{80mm} $\Bigg\downarrow H\otimes I^{\otimes n} $
\begin{align}
\Ket{\psi_1}=\frac{\Ket{0}+\Ket{1}}{\sqrt{2}}\Bigg\{\frac{\Ket{0}^{\otimes n}-\Ket{1}^{\otimes n}}{\sqrt{2}}\Bigg\}
\end{align}
\hspace{80mm} $\Bigg\downarrow U_{c_{f_{1}}} \otimes U_{c_{f_{2}}} \otimes ...\otimes U_{c_{f_{n}}} $
\begin{align}
\Ket{\psi_2}=\frac{\Ket{0}+(-1)^{f_1(0) \oplus f_1(1)}\Ket{1}}{2}\big\{\Ket{a_1 a_2 ... a_n}-\Ket{\bar{a_1}\bar{a_2}...\bar{a_n}}\big\},
\end{align}
where, $a_i \in\{0,1\}$, $U_{c_{f_{i}}}$ represents a Controlled-Unitary function applied  on the $i^{th}$ ancilla qubit. If $a_i=0$ or $1$ $\hspace{2mm}\forall i$, then $f_1(0)=f_2(0)=...=f_n(0)$ and $f_1(1)=f_2(1)=...=f_n(1)$. Hence all the functions are said to be equal. If $f_1(0)\oplus f_1(1)=0$, then $f_1(0)=f_1(1)$. On applying Hadamard gate on the first qubit one measures $\Ket{0}$, which implies that $f_1$ is constant. Since either all functions are balanced or all functions are constant, we can conclude that $f_i,\ \forall i \in [1,n]$, is constant or otherwise balanced. To summarize, 
\begin{itemize}
    \item If the first qubit is in the state $\Ket{0}$, we conclude that all the functions are constant, otherwise balanced.
    \item If all the answer qubits are correlated (either in the form of $\Ket{0}^{\otimes n}$ or $\Ket{1}^{\otimes n}$), we conclude that the functions are equal. Otherwise, at least one of them is different. 
\end{itemize}
Total number of queries required by this algorithm is $n$. Classically, we would require at least the following information to solve this task, $f_1(0),\ f_1(1),\  f_i{\oplus}f_{i+1} \ \forall i\in[1,n-1]$. This amounts to a total of $n+1$ queries. Hence, one query can be lessened by using entanglement as the main resource. It is to be noted that following a similar approach we cannot bring the number of queries down any further. In this case, each function must be queried at least once to be able to conclude whether it is constant or balanced deterministically. Since there are $n$ functions we can not have an algorithm which does the above task in less than $n$ queries. This proposal may be subject to further investigation.\par

\section{Experimental Demonstration of The Proposed Algorithm in IBM's 5-Qubit Quantum Computer \label{qal_V}}

\begin{figure}[H]
\includegraphics[scale=1]{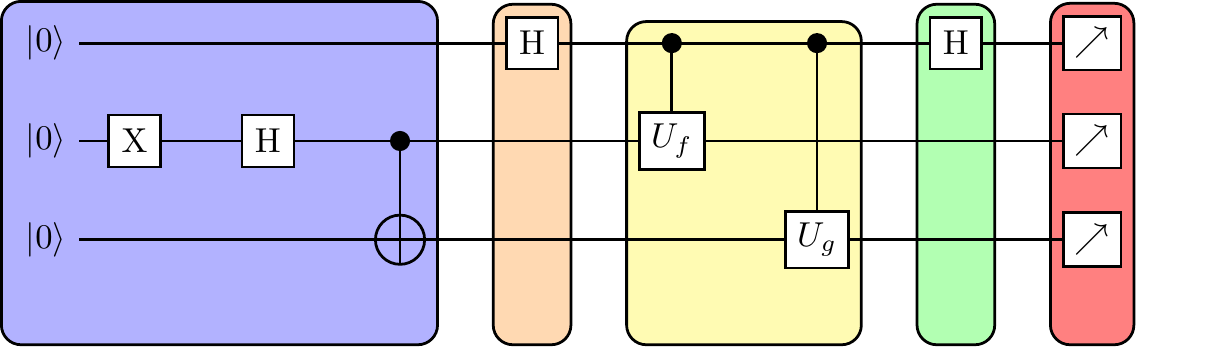}
\caption{Quantum circuit implementing the Proposed Algorithm for two functions.}
\label{qal_fig4}
\end{figure}

We consider the simplest case in which two functions are used. In Fig. \ref{qal_fig4}, the first qubit represents a register qubit while the last two qubits represent answer qubits. Here, the register qubit decides whether the functions $f$ and $g$ are constant or balanced whereas the functions are equal or unequal is decided by the answer qubits. All the qubits are initialized to $|0\rangle$ state, after applying Hadamard and on the register qubit, and entangling and applying controlled-$U_f$ and controlled-$U_g$ on the answer qubits, the composite system reads, $\frac{(|0\rangle+(-1)^{f(0) \oplus f(1)}|1\rangle)(|00\rangle-|11\rangle)}{2}$ or $\frac{(|0\rangle+(-1)^{f(0) \oplus f(1)}|1\rangle)(|01\rangle-|10\rangle)}{2}$. After applying Hadamard on the register qubit and measuring in computational basis, it is observed that for the outcome is $|0\rangle$ or $|1\rangle$ if the functions $f$ and $g$ are constant or balanced respectively. Similarly, measuring on the answer qubits, it is concluded that if the outcomes are $|00\rangle$/$|11\rangle$ or $|01\rangle$/$|10\rangle$, then the functions are equal or unequal respectively.             

The number of outcomes for the answer qubits can be reduced by following Fig. \ref{qal_fig5}, where after the controlled-functions, a CNOT, Hadamard and X gates are applied. Hence, upon the measurement on the last two qubits, there will be two outcomes, i.e, $|00\rangle$ and $|01\rangle$. It is noticed that, the functions $f$ and $g$ are equal if the outcome is $|00\rangle$ and the functions are unequal if the outcome is $|01\rangle$. 

\begin{figure}[H]
\includegraphics[scale=1]{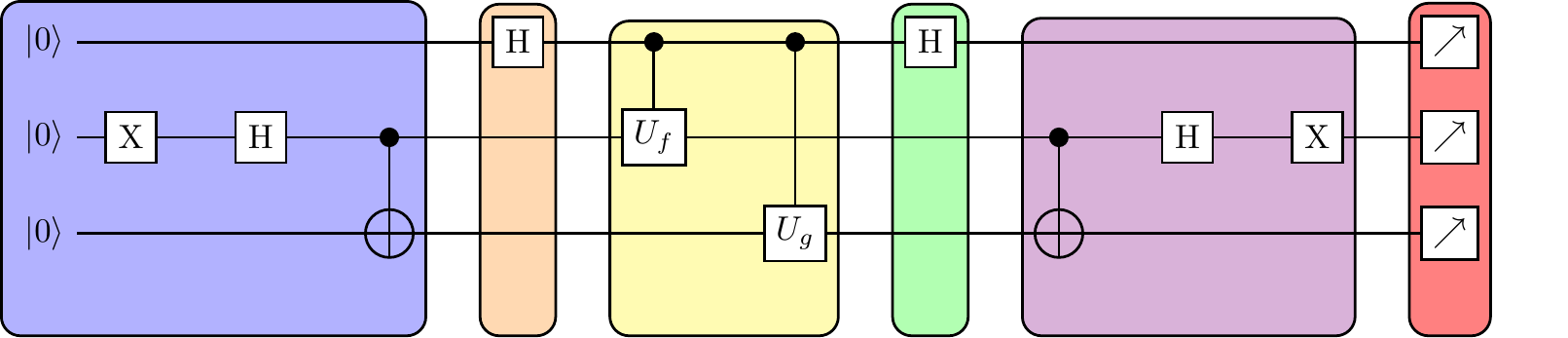}
\caption{Modified quantum circuit implementing the Proposed Algorithm for two functions for the experiment.}
\label{qal_fig5}
\end{figure}

\begin{figure}[H]
\includegraphics[scale=.9]{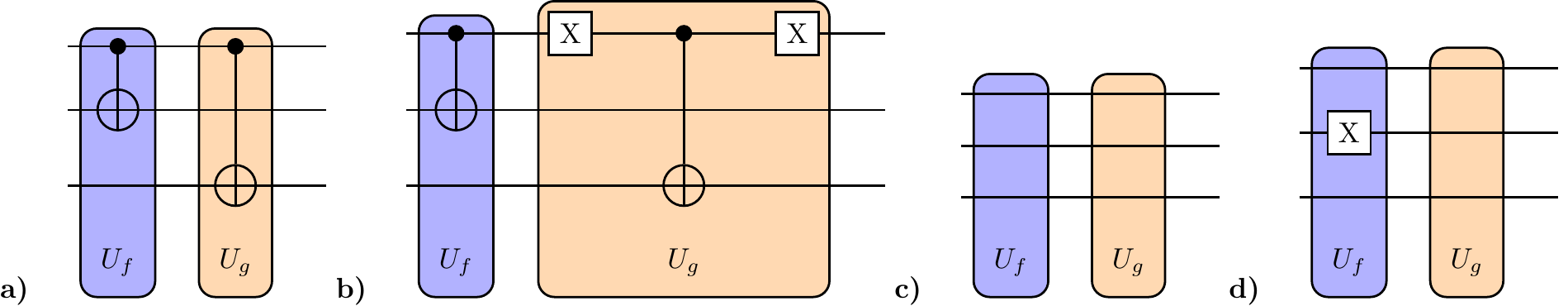}
\caption{(a) case-1: functions are `balanced and equal', (b) case-2: functions are `balanced and unequal', (c) case-3: functions are `constant and equal', (d) case-4: functions are `constant and unequal'}
\label{qal_fig6}
\end{figure}

\begin{figure}[H]
\includegraphics[scale=0.83]{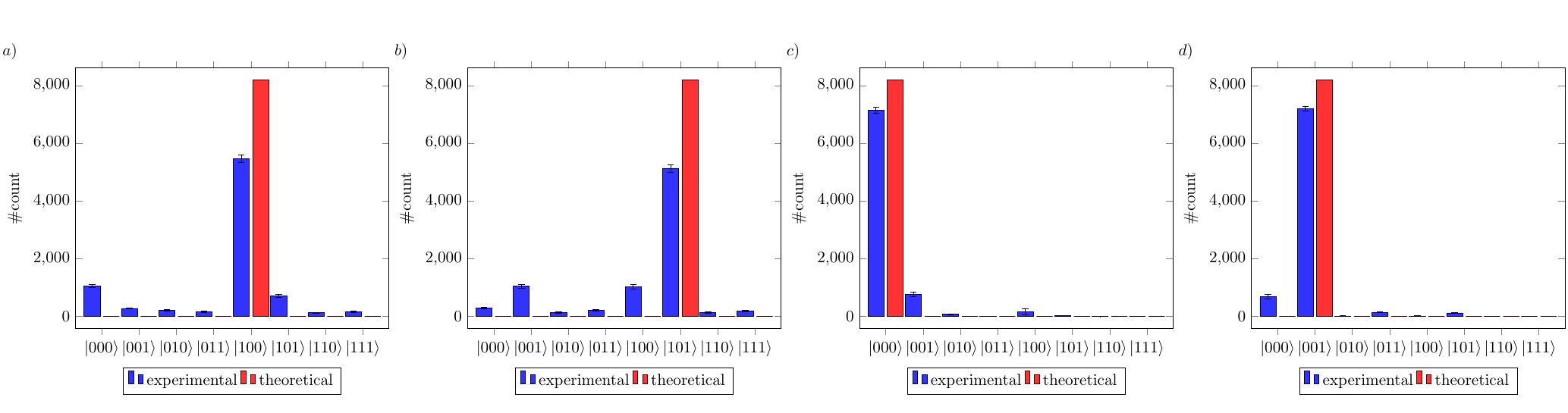}
\caption{Bar charts showing both the theoretical and experimental results. The red histograms represents the theoretical count and the blue ones counts the experimental data. (a), (b), (c) and (d) bar charts show the results for cases 1-4 respectively.}
\label{qal_fig7}
\end{figure}

Our proposed algorithm has been implemented in IBM's 5-qubit quantum computer by considering the following four possible cases. Fig. \ref{qal_fig6} illustrates the four cases; where functions are `case-1: balanced and equal', `case-2: balanced and unequal', `case-3: constant and equal', and `case-4: constant and equal'. Hence case-1, case-2, case-3 and case-4 result in the states $|100\rangle$, $|101\rangle$, $|100\rangle$ and $|111\rangle$ respectively. The quantum circuits have been designed in `ibmqx4' using the first three qubits on the chip. The experimental results have been obtained taking 10 different run results for calculating standard deviation for each case. It is to be noted that each run has been executed with 8192 number shots, where the number of shots represent the number of times we perform the measurement in the quantum circuit. The bar chart showing both the theoretical and experimental data (with standard deviation) is plotted. Fig. \ref{qal_fig7} shows the bar chart of the four cases with theoretical and experimental results. The red histograms represent the theoretical values whereas the blue ones represent experimental values. The statistical fidelities given by $F=\sum_{i=0}^{7} \sqrt{p_i^{th}p_i^{ex}}$ (where $p_i^{th}$ and $p_i^{ex}$ are the theoretical and experimental probabilities respectively for obtaining the $i^{th}$ state) \cite{qal_nil} for all the four cases have been calculated to be 0.8174, 0.7918, 0.9341 and 0.9372 respectively. It can be seen that the fidelity of case-2 is relatively lower than the other cases. It is to be noted that a relatively more number of gates are used in the case-2. As the gates and qubits in ibmqx4 have some fixed value of errors, it is expected to have a lower fidelity in the case-2. The experimental architecture of the chip ibmqx4 is given in Table \ref{qal_tab1}, where gate and readout errors, coherence and relaxation time of qubits are presented. 

\begin{table}
\centering
\begin{tabular}{ c c c c c c}
\hline
\hline
Qubits & Gate Error ($10^{-3}$) & Readout Error ($10^{-2}$) & MultiQubit Gate Error ($10^{-2}$) & $T^{||}_{1}$ ($\mu s$) & $T^{\perp}_{2}$ ($\mu s$)\\
\hline
\hline
Q0 & 0.86 & 4.60  & -    & 35.2 & 38.1 \\
Q1 & 0.69 & 5.40  & 1.99 & 57.5 & 40.5 \\
Q2 & 1.97 & 12.80 & 2-7  & 36.6 & 54.8 \\
\hline
\hline
\end{tabular}\\
$||$ Relaxation time, $\perp$ Coherence time.
\caption{The experimental architecture of the qubits.}
\label{qal_tab1}
\end{table} 

\section{Conclusion \label{qal_VI}}
We have proposed an algorithm which uses one less query than it's corresponding classical or quantum (without entanglement) algorithm which decides whether $n$  functions (given that either all the functions are constant or all are balanced) are:\ $(i)$
     equal or unequal\
    $(ii)$ constant or balanced. 

Without entanglement, it is not possible to calculate the logical functions, $f(0) \oplus f(1)$ and $f(0) \oplus g(0)$ together using a classical or a quantum computer. However, a quantum computer using entanglement as a resource is able to compute the above quantity deterministically by the help of one query to each of the functions $f$ or $g$ \cite{qal_Ashutosh}. We have made a conjecture that the number of queries cannot be reduced further using any quantum algorithm. Readers may want to further explore and try to prove or disprove it.

The proposed algorithm has been experimentally realized using IBM's 5-qubit quantum computer (ibmqx4). Four different cases for functions being `equal and balanced', `unequal and balanced', `equal and constant', and `unequal and constant' have been considered and verified experimentally. Bar charts have been plotted by counting events, and fidelity has been calculated to check the accuracy of our results. It has been found that the desired results are obtained with a high fidelity. 

\section*{Acknowledgement}
SG is financially supported by KVPY scholarship. MP and BKB acknowledge the support of INSPIRE fellowship, awarded by the Department of Science and Technology, Government of India. SG and MP would like to thank IISER Kolkata for providing hospitality during which a part of this work was completed.We are extremely grateful to IBM quantum experience project. The discussions and opinions developed in this paper are only those of the authors and do not reflect the opinions of IBM or any of it's employees.

\end{document}